\documentclass[aps,prl,twocolumn,showpacs,superscriptaddress]{revtex4}
\usepackage{graphicx}
\begin{document}


\title{$\mu$SR Studies of Magnetic-Field-Induced Effects in
High-$T_{c}$ Superconductors}


\author{A.~T.~Savici}
\author{A.~Fukaya}
\author{I.~M.~Gat-Malureanu}
\author{T.~Ito}
\author{P.~L.~Russo}
\affiliation{Dept. of Physics, Columbia University, New York, New York 10027}
\author{Y.~J.~Uemura}
\altaffiliation[author to whom correspondences should
be addressed ]{}
\affiliation{Dept. of Physics, Columbia University, New York, New York 10027}
\author{C.~R.~Wiebe}
\affiliation{Dept. of Physics, Columbia University, New York, New York 10027}
\affiliation{Dept. of Physics and Astronomy, McMaster Univ., Hamilton, ON, Canada}
\author{P.~P.~Kyriakou}
\author{G.~J.~MacDougall}
\author{M.~T.~Rovers}
\author{G.~M.~Luke}
\affiliation{Dept. of Physics and Astronomy, McMaster Univ., Hamilton, ON, Canada}
\author{K.~M.~Kojima}
\author{M.~Goto}
\author{S.~Uchida}
\affiliation{Dept. Physics, University of Tokyo, Tokyo 113-8656, Japan}
\author{R.~Kadono}
\affiliation{Institute of Materials Structure Science, KEK, Tsukuba, Japan}
\author{K.~Yamada}
\affiliation{Institute for Chemical Research, Kyoto University, Uji,
Kyoto 611-0011, Japan}
\author{S.~Tajima}
\author{T. Masui}
\affiliation{ISTEC, Shinonome 1-10-13, Koto-Ku, Tokyo 135-0062, Japan}
\author{H.~Eisaki}
\author{N.~Kaneko}
\author{M.~Greven}
\affiliation{Dept. of Applied Physics, Stanford University, Stanford, California 94305}
\author{G.~D.~Gu}
\affiliation{Brookhaven National Lab., Upton, New York 11973}
\date{\today}

\begin{abstract}
Muon spin relaxation ($\mu$SR) measurements in high transverse magnetic fields
($\parallel  \hat c$) 
revealed strong field-induced quasi-static magnetism in the underdoped
and Eu doped (La,Sr)$_{2}$CuO$_{4}$ and La$_{1.875}$Ba$_{0.125}$CuO$_{4}$, 
existing well above $T_{c}$
and $T_{N}$.  The susceptibility-counterpart of Cu spin polarization, 
derived from the muon spin relaxation rate, exhibits a divergent behavior towards
$T \sim 25$ K.  No field-induced magnetism was detected
in overdoped La$_{1.81}$Sr$_{0.19}$CuO$_{4}$, optimally doped Bi2212, and
Zn-doped YBa$_{2}$Cu$_{3}$O$_{7}$.  
\end{abstract}

\maketitle

The interplay between magnetism and superconductivity is one of
the central subjects in the study of
high T$_{c}$ superconductivity \cite{sscproceedings}.
Recently, remarkable effects
have been observed in experiments with high external
magnetic fields $B_{ext}$ applied parallel to the c-axis of single
crystal specimens.  Lake {\it et al.\/} \cite{Lake01}
found substantial increase of
inelastic neutron scattering intensity at low energy transfers
and incommensurate wavevectors
with increasing field
in optimally doped La$_{2-x}$Sr$_{x}$CuO$_{4}$ (LSCO).
More dramatic effects have been observed in
underdoped LSCO \cite{Katano,Lake02} and
in oxygen-doped La$_{2}$CuO$_{4+y}$ (LCO) \cite{Khaykovich02},
where static spin correlations are
enhanced by $B_{ext}$, as seen by
increased elastic intensities of satellite neutron Bragg peaks.
NMR experiments in high fields
in Tl$_{2}$Ba$_{2}$CuO$_{6+\delta}$ (Tl2201) \cite{Kakuyanagi03} and
YBa$_{2}$Cu$_{3}$O$_{7-\delta}$ (YBCO) \cite{Mitrovic01} detected
the presence of dynamic antiferromagnetic
spin correlations inside vortex cores.
There remain, however, many unanswered questions, such as,
(a) in which case is the field-induced magnetism static or dynamic ?
(b) does it
occur exclusively in vortex cores or everywhere in the system ?
(c) is this phenomenon generic to all the cuprates
or specific to particular series and/or doping ranges ?
(d) is the superconducting order parameter affected by the field-induced
magnetism ?

Muon spin relaxation ($\mu$SR) is a probe well suited to shed some
light on these aspects.  Following zero-field (ZF) $\mu$SR
studies on static magnetism of LSCO and YBCO superconductors \cite{Weidinger,Niedermayer}, 
our recent ZF $\mu$SR results \cite{Savici02} elucidated the
coexistence of magnetism and superconductivity in
La$_{2}$CuO$_{4.11}$ (LCO:4.11) and LSCO with $x$ = 0.12 (LSCO:0.12), where
static magnetism occurs only in a finite volume fraction,
forming nano-scale static spin stripe regions
with a radius $\sim$ 30 \AA\ (comparable to the in-plane coherence
length $\xi_{ab}$).
In $\mu$SR measurements with low transverse field (TF) ($B_{ext}$ = 0.2 T) in
La$_{2-x-y}$Eu$_{y}$Sr$_{x}$CuO$_{4}$ (LESCO) \cite{Kojima03}, we found a
tradeoff between the superfluid density and the
magnetically ordered volume fraction, which indicates that the
superconductivity and static spin ordering occur in microscopically
intertwined
but spatially separate regions.

In this paper, we report TF-$\mu$SR studies of the effect of high 
magnetic fields on several
cuprate systems, listed in Table 1, using single crystal specimens.
The values of their transition temperatures for superconducting state ($T_{c}$) and 
static magnetic order ($T_{N}$) in ZF are shown together with
the volume fraction $V_{Cu}$ of static Cu moments in ZF determined by ZF-$\mu$SR.    
Refs. \cite{Savici02,Katano,Tranquadanature} describe growth and/or characterization of 
similar (not identical) crystals with the same nominal compositions 
made by the groups who prepared the present specimens.

\squeezetable
\begin{table}
 \caption{\label{datatable}Single crystal specimens studied in this work}
 \begin{ruledtabular}
 \begin{tabular}{llcccc}
Composition & Abbreviation & T$_{c}$ (K) & T$_{N}$ (K) & V$_{Cu}$ (\%)\\
\hline
 La$_{1.88}$Sr$_{0.12}$CuO$_{4}$ & LSCO:0.12 & 28&20&40\\
 La$_{1.875}$Ba$_{0.125}$CuO$_{4}$ & LBCO &2.5&40&80-100\\
 La$_{1.75}$Eu$_{0.1}$Sr$_{0.15}$CuO$_{4}$ & LESCO &20&22&$\sim$50\\
\hline
 La$_{1.81}$Sr$_{0.19}$CuO$_{4}$ & LSCO:0.19 & 30&0&0\\
 YBa$_{2}$(Cu$_{2.979}$Zn$_{0.021}$)O$_{7}$&
 YBCO(Zn)&80&0&0\\
 (Bi,Pb)$_{2}$Sr$_{2}$CaCu$_{2}$O$_{8}$ & Bi2212 & 80 & 0 & 0\\
 \end{tabular}
 \end{ruledtabular}
 \end{table}

Specimens of
dimensions $\sim$ 8 mm x 8 mm x 1 mm are mounted in He gas-flow cryostats
with the largest face (ab-plane) perpendicular
to the muon beam direction $\hat z$,
along which
the external field was applied
with an initial muon spin polarization made perpendicular
to $\hat z$ using a spin rotator.
We used the HI-TIME $\mu$SR
spectrometer at TRIUMF to detect
high-frequency precession signals.
Time resolution and positron trajectories restricted the highest available
field to $B_{ext}$ = 6 T.

\begin{figure}[b]
\includegraphics[width=3.2in,angle=90]{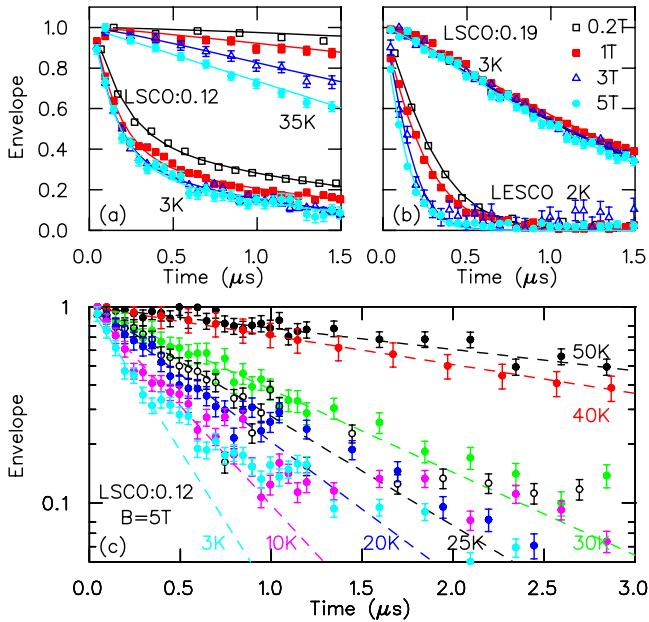}%
\caption{\label{fig1}(color)
The envelope of muon spin precession in transverse external
fields $B_{ext}$ parallel to the c-axis
observed in (a) LSCO:0.12 at $T$ = 3 K and 35 K;
(b) LSCO:0.19 at $T$ = 3 K and LESCO at $T$ = 2 K;
and (c) LSCO:0.12 in $B_{ext}$ = 5 T in a logarithmic plot
which demonstrates decay in two-component signals below $T \sim 25$ K.}
\end{figure}

Traditionally, the TF-$\mu$SR data in
high fields have been analyzed using Fourier Transforms
and/or rotating reference frame (RRF) \cite{millerprl,kadonoprb}.
Neither of them, however, clearly
display the existence of multi-component signals. As an
alternative method \cite{uemurassc79}, we extracted the amplitude
of the signal from very short time intervals ($\sim$ 50 ns) by
fitting the precession signal to a simple sinusoidal form
$A\cos(\omega t+\phi)$. In such a time domain, the amplitude and
frequency are practically constant, and the derived
amplitude $A$ for each interval represents the muon spin
relaxation function in TF, which we call the
``envelope'' of the precession signal.
Figure 1 shows the time evolution of the envelope in several systems.

A significant increase of the relaxation rate with increasing fields
is seen in LSCO:0.12 and LESCO at $T \ll T_{c}$ [Figs. 1(a) and (b); $B_{ext}$
shown in (b)], and also 
at $T > T_{c}$ and $T > T_{N}$ as shown for LSCO:0.12 in Fig. 1(a).
We obtained similar results in LBCO. 
The observed relaxation in these systems at $T < T_{N}$ is due predominantly to magnetism,
since their relaxation rates are much higher than 
those expected from the magnetic field penetration depth $\lambda$.  
In LSCO:0.12 at $T < T_{N}$, the envelope shows a clear two-component
decay, as demonstrated in the logarithmic plot of Fig. 1(c).
In contrast, essentially no field 
dependence has been observed in LSCO:0.19 (Fig. 1(b)), YBCO(Zn) and Bi2212,
in which the relaxation can be explained in terms of the 
penetration depth $\lambda$ in the superconducting state. 
A slight {\rm reduction\/} of the relaxation rate at higher fields in
YBCO(Zn) and Bi2212 can be attributed to a small increase of $B_{ext}$ relative
to $H_{c2}$ \cite{Sonier99}.    

\begin{figure}[b]
\includegraphics[width=3.2in,angle=0]{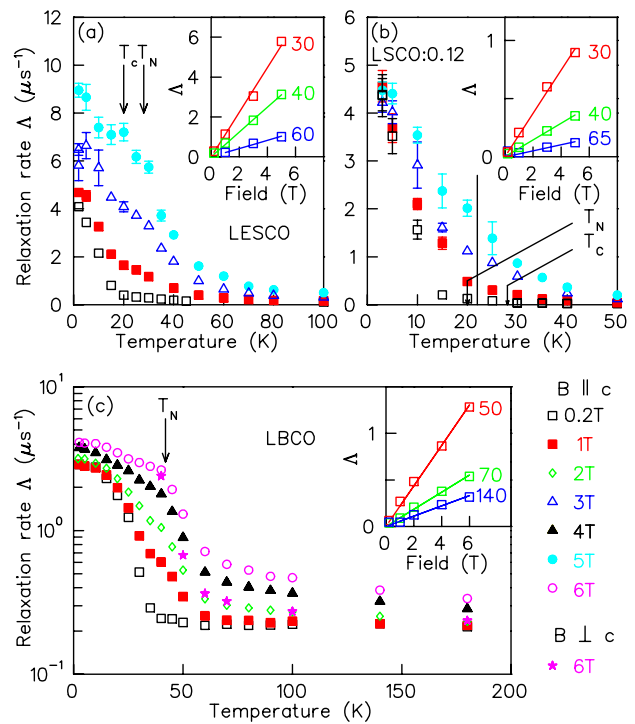}%
\caption{\label{fig2}(color) Muon spin relaxation rate $\Lambda$
in TF-$\mu$SR with $B_{ext}$ applied parallel to the c-axis
in (a) LESCO, (b) LSCO:0.12 and (c) LBCO.
The purple star represents the results in LBCO with $B_{ext} \perp \hat c$.
Points of $\Lambda$ represent the decay rate of a single-component exponential signal, except for 
$T < 25$ K in LSCO:0.12 where they represent those of the faster
component in a two-component exponential signal.
The insets show field dependence of $\Lambda$ 
at selected temperatures.}
\end{figure}

We fit the observed envelope of LBCO, LESCO, and LSCO:0.12 ($T > 25$ K)
with a single exponential decay
$\exp(-\Lambda t)$ and derived the muon spin relaxation rate
$\Lambda$, as shown in Fig. 2.  In the case of the two-component decay of LSCO:0.12 below
$T_{N}$, we fit the envelope with a sum of two exponentially decaying 
functions.  The points in Fig. 2(b) for this region
(left of the vertical solid line) correspond to the exponential
decay rates of the faster component, i.e, the slope of the broken lines in Fig. 1(c), while
the points above T = 25 K represent $\Lambda$ in the single-component fit.
These single- and two-component
exponential fits in LSCO:0.12 and LESCO are displayed by the 
solid lines in Figs. 1(a) and (b), which
exhibit a good agreement with the observed data.

In addition to large and field-dependent relaxation rate below $T_{N}$
due to local field from static Cu moments,
we see a significant field-dependent relaxation well above $T_{c}$ and $T_{N}$ 
in LESCO, LSCO:0.12 and LBCO.  As shown in the inset figures of Fig. 2,
the relaxation rates $\Lambda$ above $T_{N}$, after correction for background
relaxation due to nuclear dipolar fields and magnet-related inhomogeneous fields,
exhibit a linear relation to the external field at a given temperature.  
In LBCO, we also performed some measurements with crystals cut into 
a different orientation to have the external field perpendicular to the 
c-axis, whose results are shown by the purple stars in Fig. 2(c).
While $\Lambda$ exhibits nearly no dependence on the 
field orientation 
below $T_{N}$, we found a remarkable reduction of the field-induced effect 
above $T_{N}$ for $B_{ext} \perp \hat c$.  

From the slope of the field dependence of $\Lambda$ for $B \parallel c$-axis,
such as those shown in the insets of Fig. 2, 
we obtained $d\Lambda/dB$, which might be called 
the ``susceptibility-counterpart'' since the field-induced relaxation 
in these system is likely 
due to static random fields as discussed later.      
Figure 3(a) shows $dB/d\Lambda$, or the counterpart of inverse susceptibility,
in these three systems as a function of temperature.  
We find that $dB/d\Lambda$ shows nearly 
linear temperature dependence with the intersect to the horizontal axis at
$T \sim 25$ K common to all the three systems.  This is a behavior
analogous to inverse dc susceptibility of a {\rm ferromagnet\/} having the Curie
temperature of 25 K.   

\begin{figure}[b]
\includegraphics[width=2.3in,angle=90]{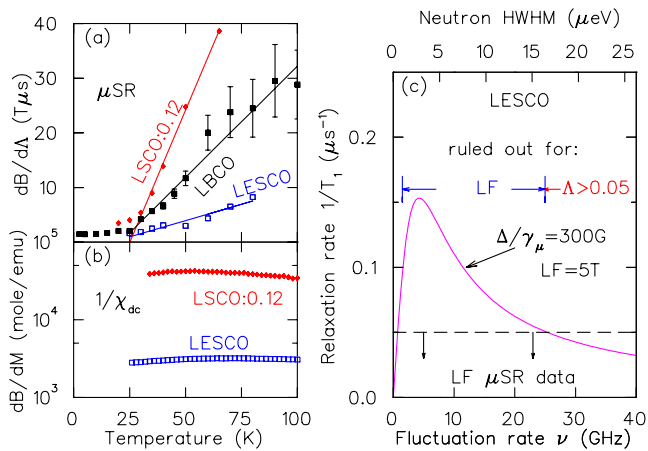}%
\caption{\label{fig3}(color) 
(a) The inverse ``susceptibility-counterpart'' $dB/d\Lambda$ derived from 
the dependence of the relaxation rate $\Lambda$ in TF-$\mu$SR on $B_{ext} \parallel \hat c$ 
= 0.2 - 6 T.
(b) The inverse of dc-susceptibility $\chi_{dc} \equiv dM/dB$ determined from 
the magnetization $M$ for $B \parallel \hat c$ = 0.2 - 5 T.
(c) Expected longitudinal relaxation rate $1/T_{1}$ in LF-$\mu$SR
for Cu spin fluctuations with the rate $\nu$ compared with the
LF-$\mu$SR results in LESCO at external field of 5 T.  $\nu$ is also displayed with
corresponding energy width (HWHM) in neutron
studies.}
\end{figure}

Using a SQUID magnetometer, we also measured 
the dc-susceptibility $\chi_{dc}$ of our specimens of LSCO:0.12
and LESCO in an external field $B \parallel \hat c$ of 0 - 5 T.  
At $T_{c} < T < 200$ K,
the magnetization $M$ exhibits linear variation with $B$, 
with the field-independent slope $dM/dB$. Figure 3(b) shows the 
inverse susceptibility $1/\chi_{dc} \equiv dB/dM$ in LSCO:0.12 and LESCO.  
The large $\chi_{dc}$ in LESCO is due to the Van Vleck term of Eu$^{3+}$.
In both systems $1/\chi_{dc}$ is nearly independent of temperature,
in a remarkable contrast to the ``counterpart from $\mu$SR'' in Fig. 3(a).

In order to study possible dynamic effects,
we have also performed $\mu$SR measurements of LESCO [and LBCO] under longitudinal
field (LF) of $B_{ext} \parallel \hat c$ = 5 T [4 T] at 
temperatures between 2 K and 150 K (of 20-30 K intervals) [2.5 and 160 K; 20 K intervals].  
We found no relaxation at any temperature, which provides 
an upper limit of the muon spin relaxation rate $1/T_{1} < 0.05 \mu$s$^{-1}$.  
When Cu moments of $\sim$ 0.5 $\mu_{B}$/Cu become static,
the internal magnetic field of 300 G would be created at the
muon site in LESCO and LBCO \cite{Savici02,Kojima03}.  For Cu spin fluctuations 
with the rate $\nu$,
we simulated $1/T_{1}$ in LF of $B_{ext} = 5$ T 
assuming a Gaussian random internal field with
the width $\Delta/\gamma_{\mu} = 300$ G,
where $\gamma_{\mu}$ denotes the gyromagnetic ratio of the muon spin.
The simulation results in Fig. 3(c) 
exceed the upperlimit of $1/T_{1}$ observed in LESCO
for $\nu$ = 1-25 $\times 10^{9}$/s, ruling out $\nu$ in this region (blue arrows in Fig. 3(c)).
When $\nu$ is significantly 
larger than $\nu \sim \gamma_{\mu}B_{ext}$ which gives the $1/T_{1}$ maximum, 
$\Lambda$ in TF should be nearly 
equal to $1/T_{1}$ in LF, contrary to the observed data
$\Lambda \gg 1/T_{1}$.  Consequently, the fast fluctuation
$\nu > 25 \times 10^{9}$ /s cannot explain observed
results.  Combined results in LF and TF now
indicate that $\nu > 10^{9}$ /s is incompatible with our data.

Thus, we consider that the observed relaxation above $T_{N}$ is due to 
quasi-static random fields induced by the external field.
Since this effect does not appear in the uniform susceptibility,
it should be related to Cu spin polarization
with finite wavevector component $q$, presumably originating from 
(possibly short ranged) stripe spin correlations
which are stabilized below $T_{N}$.
The dependence on the crystal orientation 
in LBCO rules out ``polarization of dilute magnetic impurities'' as a possible
explanation.  Good fit of the envelope to single exponential decay above $T_{N}$
suggests that the observed phenomenon is not confined to a small volume fraction
of the system.   

\begin{figure}[t]
\includegraphics[width=2.1in,angle=90]{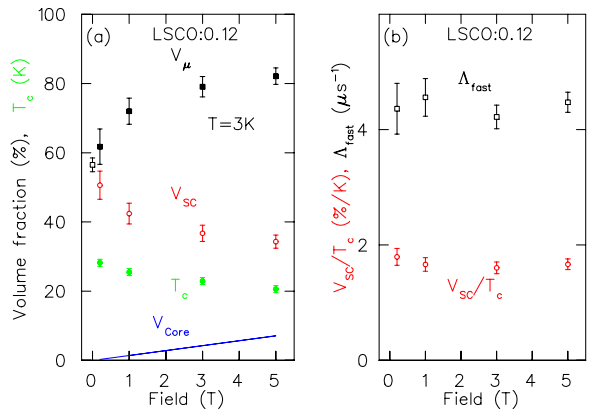}
\caption{\label{fig4}(color) (a) The volume fractions of
the fast muon decay signal $V_{\mu}$ and superconducting region $V_{sc}$
and $T_{c}$ in LSCO:0.12, and the
core region $V_{core}$ for a superconductor with $\xi_{ab} = 23$ \AA.
(b) The decay rate $\Lambda_{fast}$ at T = 3 K and $V_{sc}/T_{c}$
in LSCO:0.12.}
\end{figure}

Regarding the two component signals in LSCO:0.12 below $T_{N}$, the signal with fast (slow) 
relaxation must be due to volume fractions with (without) static magnetic order.
In Fig. 4(a), we show the amplitude fraction $V_{\mu}$ of muons involved in
the fast relaxation.  The volume fraction $V_{Cu}$ of static Cu moments is 
slightly smaller than $V_{\mu}$, since muons stopped near the edge, but outside, of the 
``magnetic islands'' are also depolarized.  After converting $V_{\mu}$
into $V_{Cu}$ using Fig. 8 of ref. \cite{Savici02}, we obtained $(1-V_{Cu}) \equiv V_{sc}$.
(Different models for the shape of magnetic regions would yield qualitatively
similar behavior of $V_{Cu}$.)
If one assumes that the slow relaxation is due to the superconducting region,
$V_{sc}$ represents the superconducting volume fraction.
Figure 4 shows that both $V_{sc}(T\rightarrow 0)$ and $T_{c}$ decreases with increasing
field, but their ratio is nearly independent of the field.  Similar proportionality
between $T_{c}$ and the superconducting volume fraction \cite{uemurareview} has been 
observed in the case of varying $T_{c}$ with (Cu,Zn) substitution in LSCO systems \cite{Nachumiprl} 
and increasing Eu concentration in LESCO \cite{Kojima03}.

In LSCO:0.12, the relaxation rate $\Lambda$ of the fast-decay component
is nearly independent of the field at $T \leq 5$ K, as shown
in Fig. 2(b) and 4(b).  This implies that the external field increases the magnetic volume
fraction but does not change the moment size.
The blue solid line $V_{core}$ in Fig. 4(a) represents volume fraction of 
the vortex core region for superconductors having the in-plane
coherence length $\xi$ = 23\AA.
We can rule out the ``vortex cores with full static magnetism''  
at $T$ = 2-3 K in YBCO:Zn, Bi2212 and LSCO:0.19, in which the observed
envelopes in $B_{ext}$ = 5-6 T indicate that
muon spins involved in such ``fast decay'' is less than 1\%\ in volume.
Static magnetism with much smaller field ($\sim$ 15 G) in the vortex core,
proposed for an underdoped YBCO \cite{millerprl}, still remains
as a possibility in these three systems.
  
The strong field dependent effects above and below $T_{N}$ were observed
only in the systems which 
exhibit static magnetism in ZF in partial or neally full
volume fraction $V_{Cu}$ (see Table 1.).  Thus, 
the field-induced relaxation is not generic to all the cuprate
systems but is confined to systems having competing magnetic
state very close in free energy to their superconducting state.  
The stronger effect for the field applied perpendicular
to the CuO$_{2}$ planes in LBCO suggests that this phenomenon 
could be related to suppression of superconductivity
caused by the field.  Though the superconducting 
$T_{c}$'s are rather low in underdoped LSCO:0.12, LBCO and LESCO
systems, the ``dynamic superconductivity''\cite{uemurareview,Anderson} detected by the 
Nernst effect \cite{Wang} might extend up to $T \sim$ 150 K 
(``Nernst region'').
The suppression of dynamic superconductivity could favor
competing antiferromagnetism \cite{Sachdev}.  

Recently, a very small diamagnetic magnetization $M_{dia}$ was detected in Bi2212 in 
the Nernst region above $T_{c}$ \cite{Ongdiamag}.
The field and temperature dependences of $M_{dia}$ follow behavior similar to 
that of $\Lambda$ in $\mu$SR.
This suggests a possibility that the observed 
relaxation above $T_{c}$ is due to the dynamic
supercurrent screening $B_{ext}$.
However, if we assume $\Lambda = \alpha M_{dia}$ and obtain the proportionality
factor $\alpha$ using the results in Bi2212 near T = 0, that factor
gives the value of $\Lambda$ more than 10 times smaller than the observed value for
a reasonable estimate of $M_{dia}$ above $T_{c}$ 
in LSCO and LESCO using the Nernst results.  
Furthermore, it is difficult \cite{Anderson} to expect the dynamic diamagnetism 
to have the time scale slower than
$t \sim 1\mu$s, required for a flux vortex lattice to produce field inhomogeneities observable
in $\mu$SR.
These difficulties have to be resolved before the dynamic screening scenario is adopted.

In conclusion, our $\mu$SR measurements in high TF revealed field-induced
magnetism in LSCO:0.12, LESCO, and LBCO existing in a wide
range of temperature above and below $T_{c}$ and $T_{N}$.  This phenomenon
is not common to all the cuprate superconductors, but rather confined
to those systems having magnetically ordered state closely competing
with the superconducting state.

We acknowledge financial supports from
NSF DMR-01-02752 and CHE-01-11752 at Columbia,
DOE DE-FG03-99ER45773 and DE-AC03-76SF00515 at Stanford U.,
DOE DE-AC02-98CH10866 at BNL,
NSERC and CIAR (Canada) at McMaster,
NEDO (Japan) at ISTEC and U. Tokyo, and MEXT (Monkashyo) (Japan)
Scientific Research (B) and (S) at
Kyoto U. and U. Tokyo.

\end{document}